\documentclass[iop,apj]{emulateapj}
\usepackage{booktabs}

\usepackage{graphicx,epstopdf,hyperref,color,multirow,amsmath,booktabs}

\newcommand{\Teff}{\mbox{$T_{\mathrm{eff}}$}}
\newcommand{\EBR}{\mbox{$E$(BP $-$ RP)}}

\begin{document}
\title{Machine Learning Regression of extinction in the second $Gaia$ Data Release }

   \author{Yu Bai\altaffilmark{1}}
   \author{JiFeng Liu\altaffilmark{1,2}}
   \author{YiLun Wang\altaffilmark{1,2}}
   \author{Song Wang\altaffilmark{1}}

\affiliation{$^1$ Key Laboratory of Optical Astronomy, National Astronomical Observatories, Chinese Academy of Sciences,
       20A Datun Road, Chaoyang Distict, Beijing 100012, China; ybai@nao.cas.cn}
\affiliation{$^2$ College of Astronomy and Space Sciences, University of Chinese Academy of Sciences, Beijing 100049, China}
\affiliation{$^3$ National Astronomical Observatories, Chinese Academy of Sciences,
       20A Datun Road, Chaoyang Distict, Beijing 100012, China}

\begin{abstract}
Machine learning has become a popular tool to help us make better decisions and predictions, based on experiences, observations
and analysing patterns within a given data set without explicitly functions. In this paper, we describe an application of
the supervised machine-learning algorithm to the extinction regression for the second $Gaia$ data release, based on the combination
of Large Sky Area Multi-Object Fiber Spectroscopic Telescope, Sloan Extension for Galactic Understanding and Exploration and
the Apache Point Observatory Galactic Evolution Experiment. The derived extinction in our training sample is consistent with other
spectrum-based estimates, and its standard deviation of the cross validations is 0.0127 mag.
A blind test is carried out using the RAdial Velocity Experiment catalog, and the standard deviation is 0.0372 mag. Such precise
training sample enable us to regress the extinction, {\EBR}, for 133 million stars in the second $Gaia$ data release.
Of these, 106 million stars have the uncertainties less than 0.1 mag, which suffer less bias from the external regression.
We also find that there are high deviations between the extinctions form photometry-based methods, and
between spectrum- and photometry-based methods. This implies that spectrum-based method could bring
more signal to a regressing model than multi-band photometry, and a higher signal-to-noise ratio would acquire a more reliable
result.

\end{abstract}

\keywords{stars: fundamental parameters --- methods: data analysis --- techniques: spectroscopic }

\section{Introduction}
\label{sec:intro}
Machine learning has been a dominant force in today$'$s world and widely used across a variety of domains,
owing to its incredibly powerful ability to
make predictions or calculated suggestions for large amounts of data. In the domains of modern astronomy,
high dimensional data consisted of billions of sources become available in recent years, which expand our understanding
of the Milky Way to a new frontier. However, an obstacle to such understanding is thick layers of dust in major
parts of our Galaxy. Thanks to dedicated large photometric, astrometric, and spectroscopic surveys, we are
now able to map the Milky Way in a much more accurate fashion.

One of the most ambitious survey is the European Space Agency mission $Gaia$ \citep{Gaia16}, which is performing an
all-sky astrometric, photometric and radial velocity survey at optical wavelength. The primary objective of the $Gaia$
mission is to survey more than one billion stars, in order to investigate the structure, the origin and subsequent
evolution of our Galaxy. The recent $Gaia$ Data Release 2 ($Gaia$ DR2; \citealt{Gaia18}), covered the first 22 months
of observations with $G$-band photometry for a total of 1.69 billion sources. Of these, 1.38 billion sources also have
the integrated fluxes from the BP and RP spectrophotometers, which span 3300$-$6800 {\AA} and 6400$-$10500 {\AA}, respectively.

These three broad photometric bands have been used to infer astrophysical parameters for about 10$^8$ stars \citep{Andrae18}.
A machine learning algorithm, random forest (RF), has been applied to regress stellar effective temperatures ({\Teff}).
Using in addition the parallaxes, they have estimated the line-of-sight extinction. The accuracy of the {\Teff} suffers
small size of training sample \citep{Pelisoli19,Sahlholdt19,Bai19b}, and would further bias the extinction estimation.

In order to present unbiased extinction, we require larger amounts of data with higher accuracy.
The availability of spectrum-based stellar parameters for large numbers is now possible thanks to the observations
of large Galactic spectral surveys.
Large Sky Area Multi-Object Fiber Spectroscopic Telescope (LAMOST; \citealt{Luo15}) data release 5 (DR5)
was available in December of 2017, which includes over eight millions observations of stars \footnote{See http://dr5.lamost.org/.}.
One of the catalog mounted on the archive is A-, F-, G- and K-type stars catalog, in which the stellar parameters, {\Teff},
log$g$ and [Fe/H] are determined by the LAMOST stellar parameter pipeline \citep{Wu14}. This archive data after six
years' accumulation is a treasure for various studies, especially for machine learning, since it largely enriches the diversity
of training samples \citep{Bai19a}. Diversity of a sample in a parameter space has been proved to be an influential aspect, and has strong impact on
overall performance of machine learning (\citealt{Wang09a}, \citealt{Wang09b}).

The large amount of such spectroscopic data provides us an opportunity to apply machine learning technology
to regress the line-of-sight extinction effectively.
In Section \ref{sec:meth}, we present validation samples and a method of the extinction prediction with the synthetic photometry. The algorithm
and the blind test are also
described in the section. We apply the regressor and present a revised
version of {\EBR} catalog for $Gaia$ DR2 in Section \ref{sec:res}. In Section \ref{sec:sum}, we discuss the comparisons with the extinction
and its coefficients from other studies.

\section{Methodology}
\label{sec:meth}

\subsection{Observational Data}
The A-, F-, G- and K-type stars catalog of LAMOST DR5 includes the estimates of the stellar parameters with the application of a
correlation function interpolation \citep{Du12} and Universit\'{e} de Lyon spectroscopic analysis software \citep{Koleva09}.
These two approaches are based on distribution and morphology of absorption lines in normalized stellar spectra, independent
from Galactic extinction.  The standard deviations of {\Teff}, log$g$ and [Fe/H] are $\sim$110 K, 0.19 dex and 0.11 dex, respectively
\citep{Gao15}. We extract 4,340,931 unique stars in the catalog, and cross match them to $Gaia$ DR2 with a radius of 2 arcseconds,
which yields 4,249,013 stars.

We also take advantage of the stellar parameters in Sloan Extension for Galactic Understanding and Exploration
(SEGUE; \citealt{Yanny09}). The spectra are processed through the SEGUE Stellar Parameter Pipeline
(SSPP; \citealt{Allende08,Lee08a}; \citealt{Lee08b,Smolinski11}), which uses a number of methods to derive accurate
estimates of stellar parameters, {\Teff}, log$g$, [Fe/H], [$\alpha$/Fe] and [C/Fe]. The typical uncertainties are 130 K,
0.21 dex and 0.11 dex for {\Teff}, log$g$ and [Fe/H], respectively \citep{Allende08}. We perform a cross-match with $Gaia$ DR2,
and obtain 1,037,433 stars.

Different from the upper two surveys that are in optical band, the Apache Point Observatory Galactic Evolution Experiment (APOGEE),
as one of the programs in both SDSS-III and SDSS-IV, has collected high-resolution ($R$ $\sim$ 22,500) high signal-to-noise (S/N $>$ 100)
near-infrared (1.51$-$1.71 $\mu$m) spectra of 277,000 stars (data release 14) across the Milky Way \citep{Majewski17}.
These stars are dominated by red giants selected from the Two Micron All Sky Survey.
Their stellar parameters and chemical abundances are estimated by the APOGEE Stellar Parameters and Chemical
Abundances Pipeline (ASPCAP; \citealt{Meszaros13,Garcia16}). The {\Teff}, log$g$ and [Fe/H] precise to 2\%, 0.1 dex, and
0.05 dex, respectively. We cross-match these stars with $Gaia$ DR2, and obtain 275,019 stars.

We here only adopt data from spectroscopic surveys, since their stellar parameters are highly
reliable \citep{Mathur17}, compared to photometric catalogs, e.g., Kepler Input Catalog. As a result, there are 5,561,465
$Gaia$ matched stars. We then use the criteria in \citet{Bai19b} to select the stars with good photometry,
and there are 3,558,618 stars left in our training sample.

\begin{figure}
   \centering
   \includegraphics[width=0.5\textwidth]{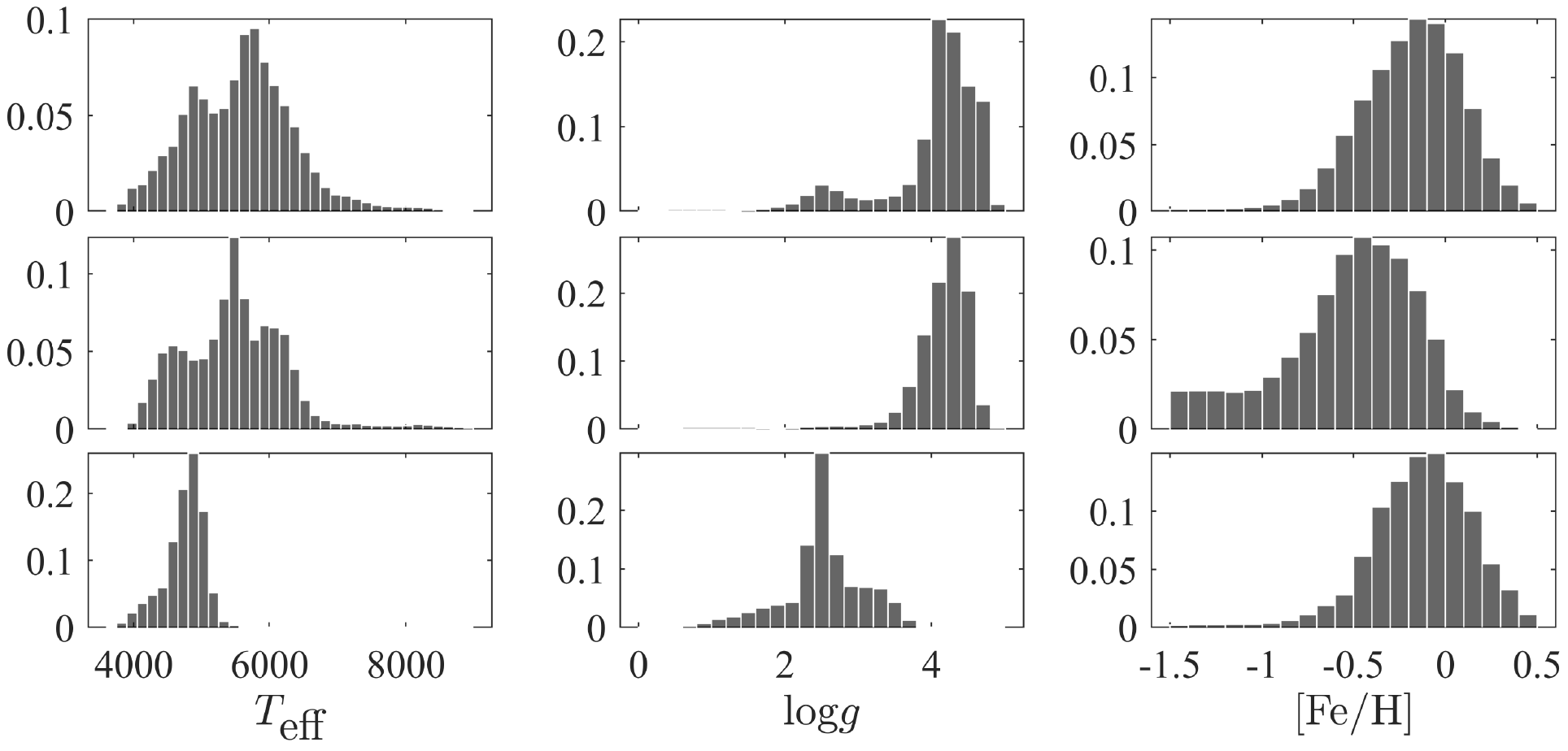}
   \caption{Stellar parameters distributions. Upper panels: LAMOST parameters. Middle panels: SSPP parameters. Lower panels: APOGEE parameters.
   \label{Par} }
\end{figure}

\begin{figure}
   \centering
   \includegraphics[width=0.5\textwidth]{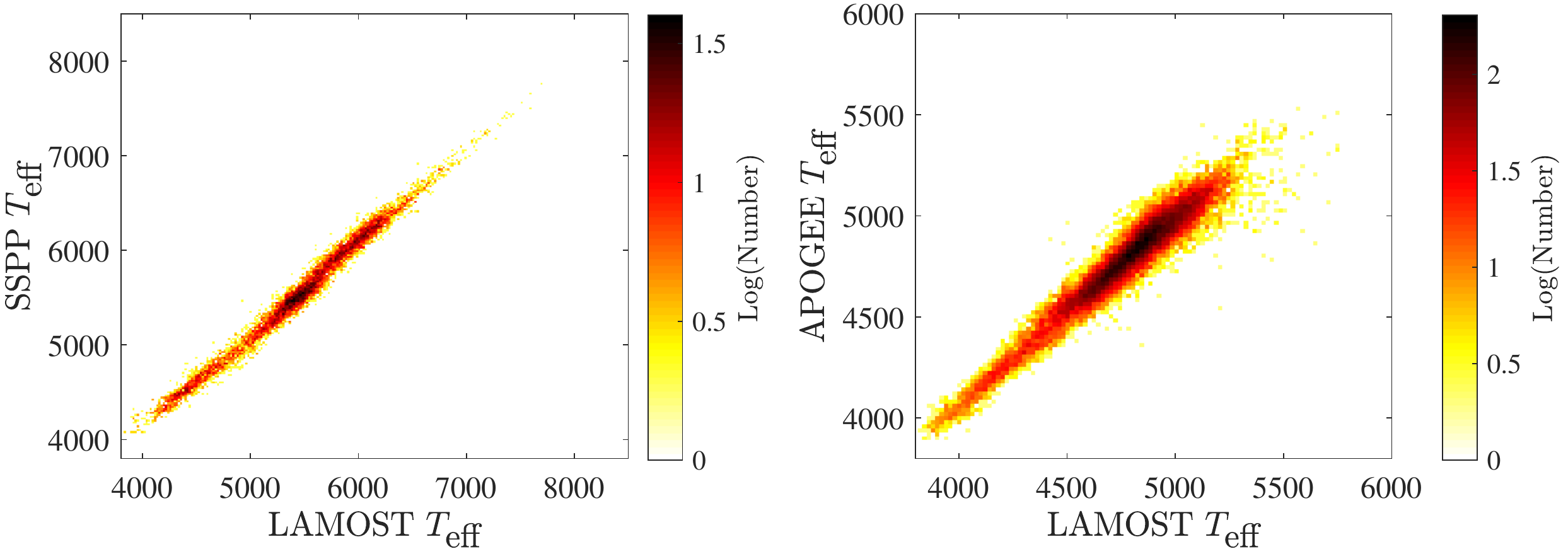}
   \caption{One-to-one correlations for the overlap stars. Left panel: the correlation between the LAMOST and SSPP {\Teff} in our training sample.
   Right panel: the correlation between the LAMOST and APOGEE {\Teff} in our training sample.
   \label{OneOne_Tr} }
\end{figure}

The stellar parameters distributions are shown in Figure \ref{Par}. The training sample is dominated by F, G, and K stars with solar like abundance.
The stars in APOGEE are mainly giants, while most of the stars in LAMOST and SSPP belong to main sequence. The RAdial Velocity Experiment (RAVE) isn't included in the
training sample, and we apply the RAVE stars to the blind test in Section \ref{sec:blin}.

We check the overlaps between LAMOST, SSPP and APOGEE, and present the one-to-one correlations of the stellar temperatures in Figure \ref{OneOne_Tr}.
There are deviations among three catalogs, which is mainly due to the difference of the pipelines \citep{Luo15}. Such systematic uncertainties are present in
Section \ref{sec:blin}.
We here don't select or remove these overlap stars or the stars that observed multiple times.
These stars share equal weight in our regression, and the deviations among catalogs or among observations are going to be propagated
to the uncertainties of results.

\subsection{Synthetic Photometry}
In order to derive extinction for the training stars, we use the BT-Dusty grid
(\citealt{Allard09,Allard11}, \citealt{Allard12})\footnote{https://phoenix.ens-lyon.fr/Grids/BT-Dusty/} of the PHOENIX
photospheric model at Theoretical Model Services (TMS)\footnote{http://svo2.cab.inta-csic.es/theory/main/} to calculate a synthetic
color, BP $-$ RP, and compare it to the color in $Gaia$ DR2. The synthetic color depends on three stellar parameters, {\Teff},
log$g$ and [Fe/H], which is different from the temperature-dependent color used in \citet{Andrae18}.
We adopt the transmission curves of $Gaia$ DR2 passbands \footnote{https://www.cosmos.esa.int/web/gaia/iow\_20180316/}. Different curves
would result in different colors and introduce uncertainties to the results \citep{Maiz18}, but such difference is unobvious, about some mmag.

\begin{figure}
   \centering
   \includegraphics[width=0.5\textwidth]{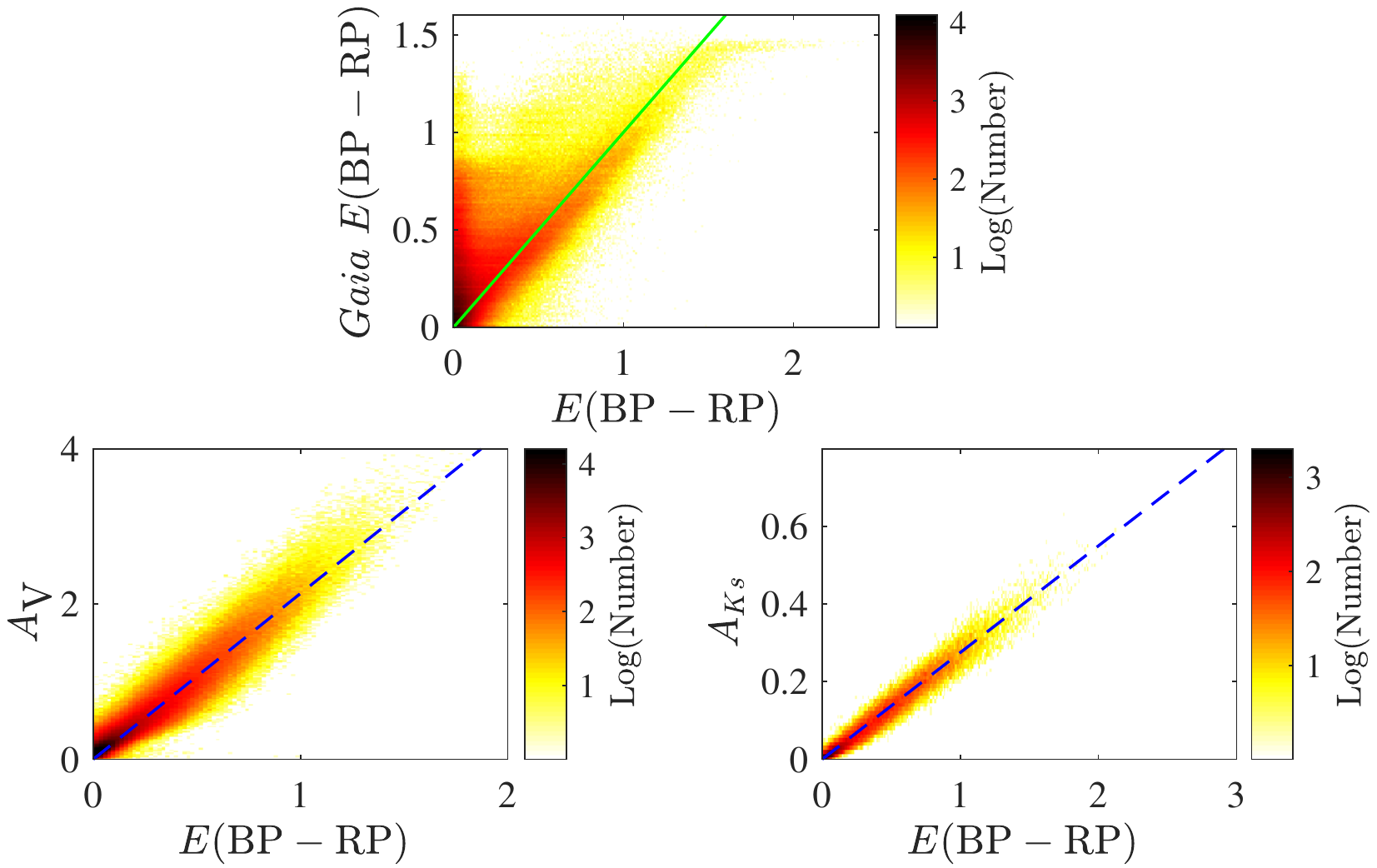}
   \caption{One-to-one extinction correlations. Upper panel: the correlation between the {\EBR} in $Gaia$ DR2 and in our training sample.
   Lower panels: LAMOST $A_V$ (left) and APOGEE $A_{Ks}$ (right) estimated by Bayesian methods vs. the {\EBR}
   in our training sample. The best linear fits are shown as the blue dashed lines. The color bars are the density of the stars in the logarithmic scale.
   \label{EBV}}
\end{figure}

We present the one-to-one correlations between the {\EBR} in $Gaia$ DR2 and those derived from the spectrum-based results in Figure \ref{EBV}.
In the upper panel, the outliers remain at {\EBR} $\sim$ 0 with $Gaia$ {\EBR} $>$ 1.5 is due to the outlier filtration \citep{Andrae18,Arenou18}.
Except these outliers, there are still many stars with the extinction overestimated by $Gaia$ DR2. It is expected, since
the {\Teff} is underestimated by $Gaia$ (Figure 3 in \citealt{Bai19b}). A lower temperature would result in higher extinction for the same sample.

A novel Bayesian method developed by \citet{Pont04} and \citet{Binney14} has been used for stars in the LAMOST survey \citep{Wang16a},
which has demonstrated the ability to obtain accurate distance and extinction. There are 1,062,590 cross-matched stars with valid extinction in their catalog.
The one-to-one correlation is shown in the lower-left panel of Figure \ref{EBV}. The best linear fit is $A_V$ = (2.138$\pm$0.001) $\times$ {\EBR}.
\citet{Wang16b} has applied a similar method on APOGEE stars to estimate their distance and extinction. We cross match these stars
with our training sample, and there are 65,471 stars left. The best fit is $A_{Ks}$ = (0.2752$\pm$0.0003) $\times$ {\EBR}.
These two good linear relations indicate that our extinction is consistent with other spectrum-based results.

\subsection{Algorithm}

The bagged regression tree of RF algorithm \citep{Breiman01} is adopted to build the regressor.
In brief, the working theory of the RF is that it builds an ensemble of unpruned decision trees and merges them together to obtain a
more accurate and stable prediction. One big advantage of RF is fast learning from a very large number of data. This algorithm has been widely used
for classification, while the RF regression isn't popular. An important example of RF regression is \citet{Miller15}, and one of the
best introduction of RF is \citet{Hastie09}.

We add two additional parameters, temperatures and their uncertainties given by \citet{Bai19b}, to the combination of the input:
{\Teff}, {$\Delta$\Teff}, $l$, $b$, $\varpi$, $\Delta\varpi$, $\mu_{\alpha}$, $\mu_{\delta}$, BP $-$ $G$ and $G$ $-$ RP.
Such combinations has the best performance on the {\Teff} regression, and it would be the best way to decouple
the extinction from the temperatures.

Then, we apply the 20 folded cross validations to test the performance of the regression.
The cross validation partitions the sample into twenty randomly chosen folds of roughly equal size. One fold is used to validate
the regression that is trained using the remaining folds.
This process is repeated twenty times such that each fold is used exactly once for validation. The 20 folded cross validation
can provide an overall assessment of the regression.

\begin{figure}
   \centering
   \includegraphics[width=0.5\textwidth]{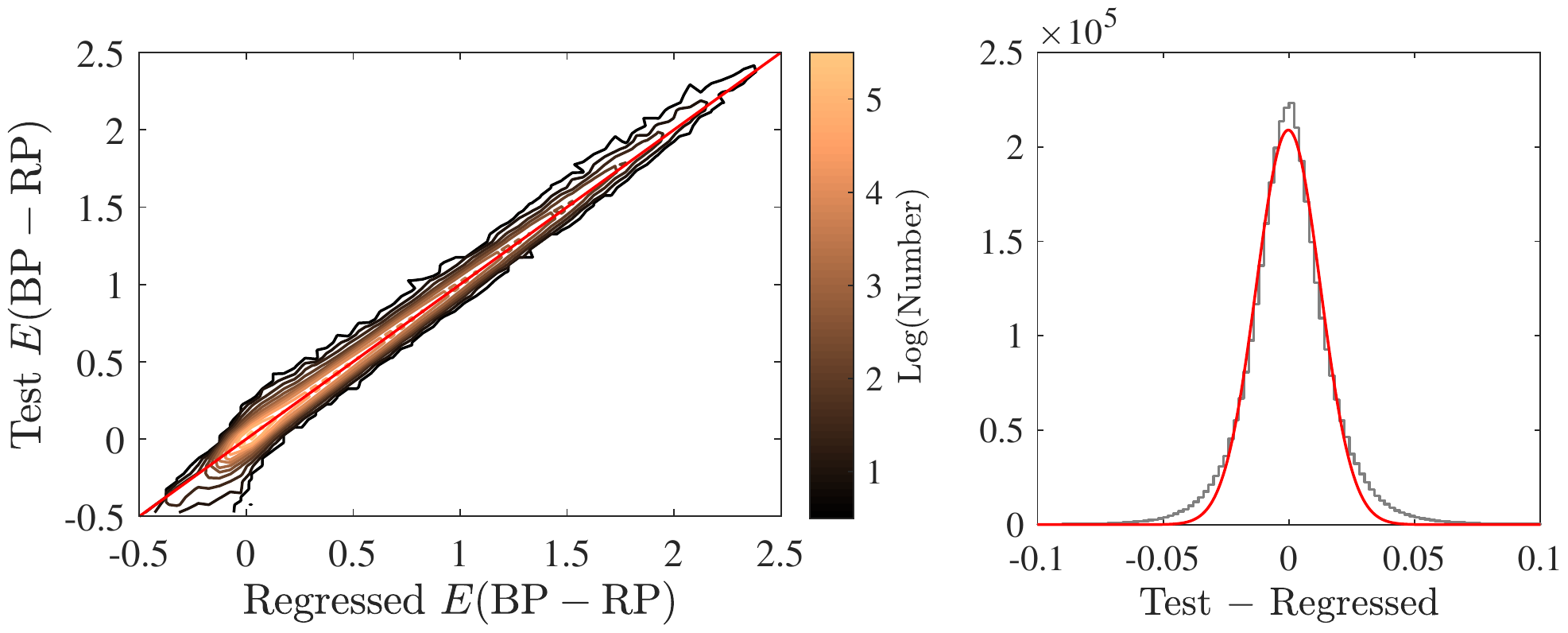}
   \caption{The results of the cross validations. Left panel: one-to-one correlation of the cross validation . The color bar is
            the density contour in the logarithmic scale. Gaussian fit (red) of the total residual
            (black) is shown in the right panel.
   \label{CV}}
\end{figure}

The one-to-one correlation of the cross validations is shown in the left panel in Figure \ref{CV}.
The Gaussian fit to the total residuals is shown in the right panel, and the fitted offset ($\mu$) and the standard deviation ($\sigma$) are
listed in Table \ref{Table1}.

\begin{figure}
   \centering
   \includegraphics[width=0.5\textwidth]{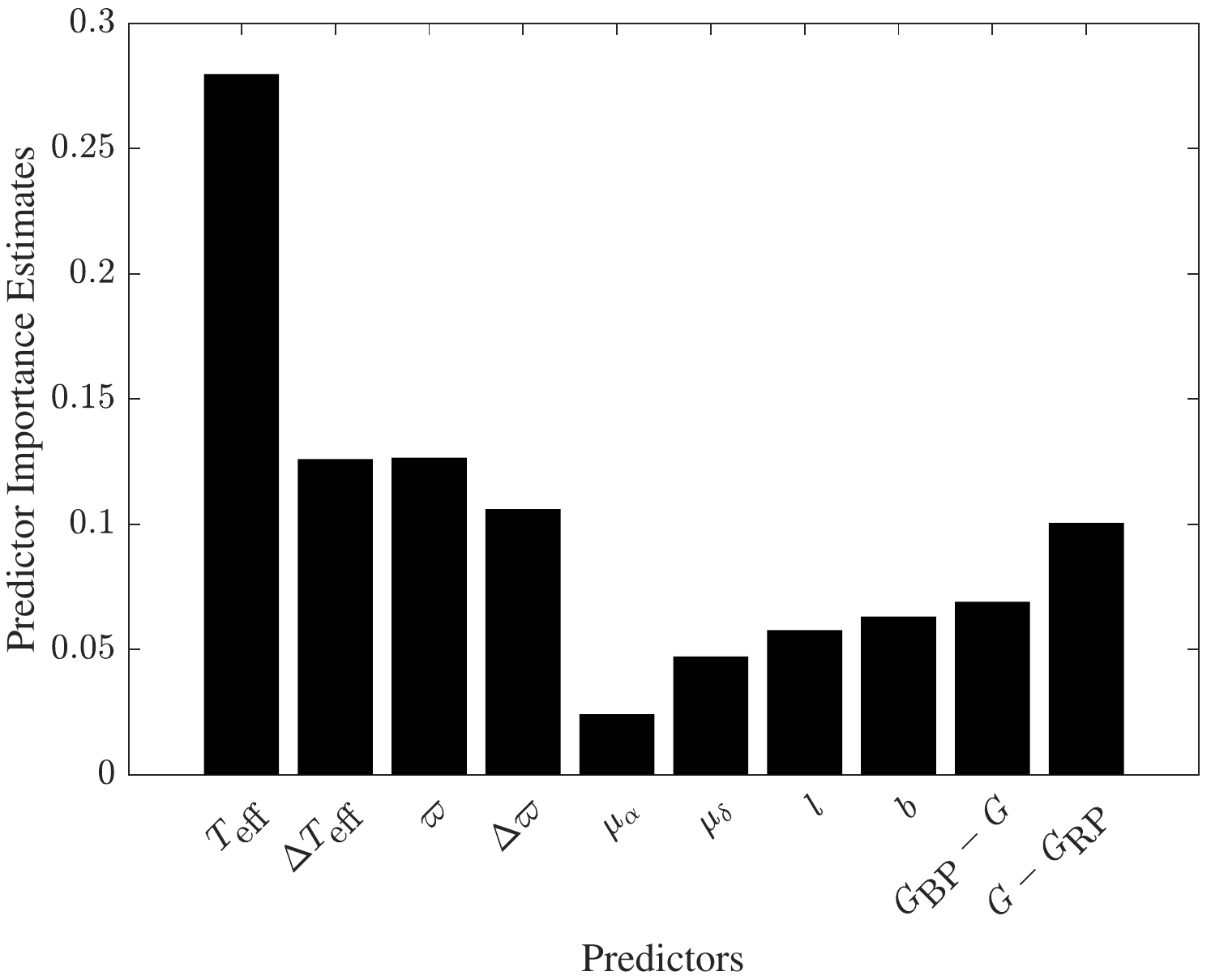}
   \caption{Important estimates of the regressor: stellar effective temperature, parallax and its error, proper motions, Galactic position,
   and two $Gaia$ colors.
   \label{Imp}}
\end{figure}

The importance estimates of the regression are shown in Figure \ref{Imp}. The temperature becomes the most important parameter,
while other parameters have similar importance. This proves that it is effective to add {\Teff} to the combination of the input parameters.
The importance of proper motions is lower than those of the $Gaia$ colors, which is different to the results in \citet{Bai19b}.
This implies that its less relevant than colors in our extinction regressing process.

\subsection{Blind Tests}
\label{sec:blin}

An independent blind test is effective technology to avoid systematic flaws, such as poor construction of training/test splits,
inappropriate model complexity, and misleading test metrics \citep{Bai19a,Guyon19}. It evaluates the prediction accuracy with
data that are not in the training sample, and provides validation that a regressor is working sufficiently to output reliable results.

The RAVE is designed to provide stellar parameters to complement missions that focus on obtaining
radial velocities to study the motions of stars in the Milky Way¡¯s thin and thick disk and stellar halo \citep{Steinmetz06}.
Its pipeline processes the RAVE spectra and derives estimates of \Teff, log $g$, and [Fe/H] \citep{Kunder17}. Using these
parameters, \citet{Binney14} applied a Bayesian method to estimate the interstellar extinction with the uncertainties $A_\textrm{V}$
$\sim$ 0.1 mag. We cross match the catalog with $Gaia$ DR2, which yields 192,483 stars.

\begin{figure}
   \centering
   \includegraphics[width=0.5\textwidth]{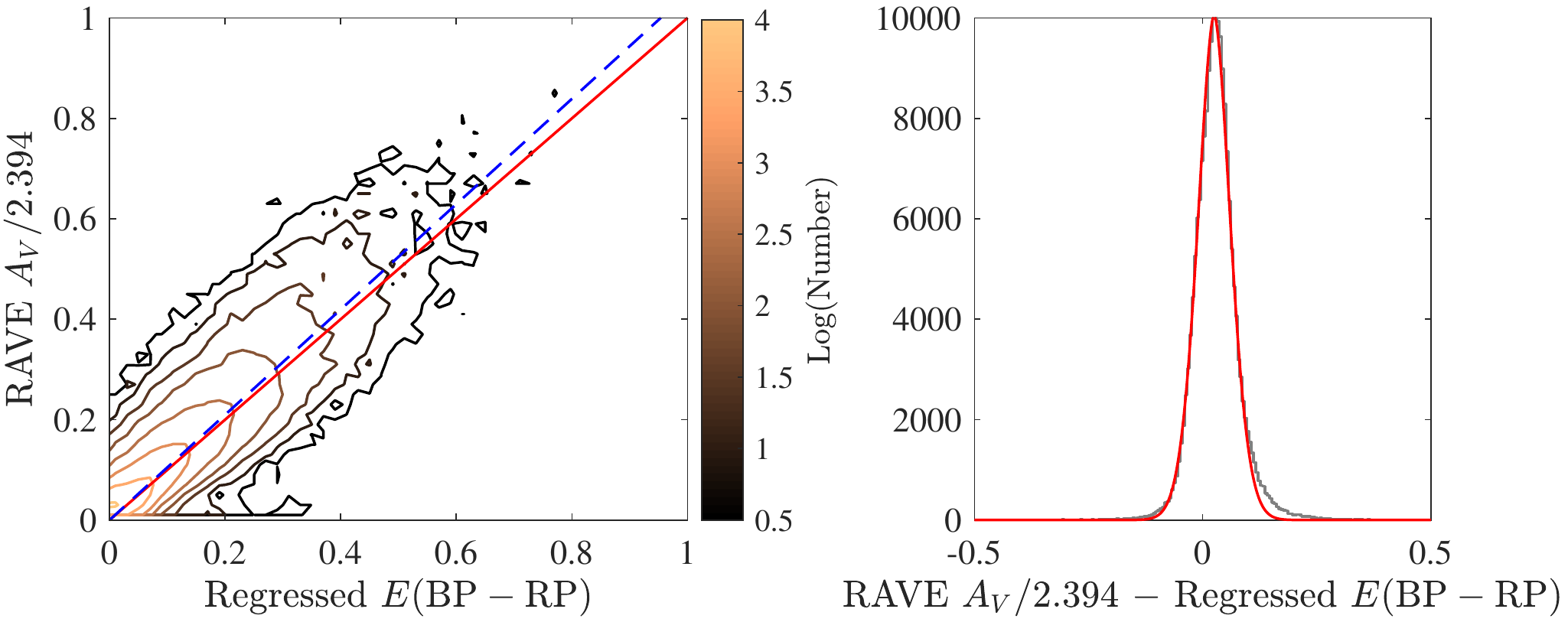}
   \caption{The blind test results. One-to-one correlation between the RAVE extinction and the regressed extinction is shown in the left panel.
            The coefficient of 2.394 \citep{Wang19} is adopted to convert $A_V$ to {\EBR}. The best linear fit is shown as a blue dashed line.
            Gaussian fit (red) of the total residual (black) is shown in the right panel.
   \label{rave}}
\end{figure}

We here adopt the extinction coefficient value, 2.394 in \citet{Wang19}, to convert RAVE $A_\textrm{V}$ to {\EBR}, and the one-to-one correlation
is shown in Figure \ref{rave}. The fitted slope is close to one, 1.044 $\pm$ 0.002, and Table \ref{Table1} lists the parameters of
the Gaussian fit to the total residuals. These imply that our regressor is reliable, and it can determine {\EBR} with fair accuracy.
It should been noted that the extinction conversion in broad-band filters depends not only on the extinction law, but also on {\Teff},
and extinction itself \citep{Girardi08}. However, this topic is beyond the main result of this paper, and we just adopt the latest coefficient
value to make a bind test.

\begin{figure}
   \centering
   \includegraphics[width=0.5\textwidth]{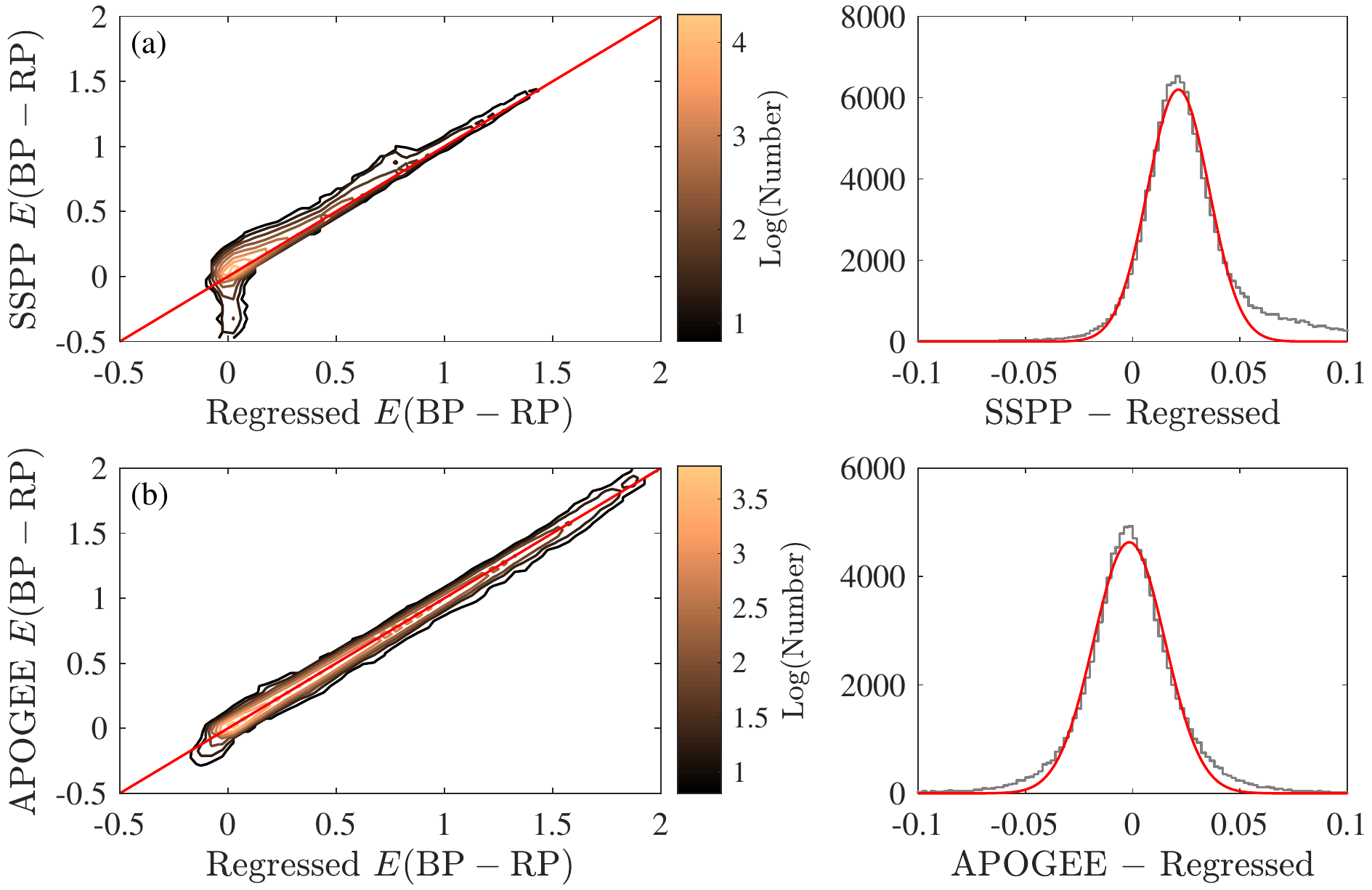}
   \caption{Density contours of one-to-one correlations (left column) and
            Gaussian fits of the total residual (right column). Two catalogs are used for
            training and the third one for a blind test. The test catalogs are: (a) SSPP,
            (b) APOGEE.
   \label{BlindT}}
\end{figure}

\citet{Bai19b} applied the subregressors to test the accuracy of the regressor, since all the spectrum-based catalogs were used
for training. The subregressors could also test the systematic uncertainty among different surveys. We here train the subregressors with
two catalogs and use the third one to test these subregressors. The LAMOST DR5 is always included in the training set, since it accounts
for 93\% of the stars in our training set. We present the results of the tests in Figure \ref{BlindT}, and list the parameters of the
Gaussian fit to the total residuals in Table \ref{Table1}. It shows that the offsets are below 0.022 mag and standard deviations are less
than 0.017 mag, which is consistent with the results of other spectrum-based methods (\citealt{Wang16a}, \citealt{Wang16b}).

\begin{deluxetable}{lrrc}
\tablecaption{Results of cross validations and blind tests \label{Table1}}
\tablehead{& $\mu$ & $\sigma$ & RMSE
           }
\startdata
Cross Validation       & $-$0.3  $\pm$ 0.2     &  12.7 $\pm$ 0.2 & 18\\
SSPP                   &    21.3  $\pm$ 0.4    &  14.3 $\pm$ 0.4 & 47\\
APOGEE                 & $-$1.5  $\pm$ 0.3     &  16.7 $\pm$ 0.3 & 31\\
\midrule
RAVE                   &    25.1 $\pm$ 0.4     &  37.2 $\pm$ 0.4 & 58
\enddata
\tablecomments{The unit is 10$^{-3}$ mag.
}
\end{deluxetable}

\section{Result}
\label{sec:res}

\begin{deluxetable}{cl}
\tablecaption{Results of our regression for $Gaia$ DR2 \label{Table2}}
\tablehead{ Source ID & Regressed {\EBR}
           }
\startdata
2448780173659609728 & 2.05 $\pm$ 0.28\\
2448781208748235648 & 0.034 $\pm$ 0.014\\
2448689605685695488 & 0.015 $\pm$ 0.019\\
2448689777484387072 & 0.490 $\pm$ 0.118\\
2448783991887042176 & 0.095 $\pm$ 0.037\\
2448690258520723712 & 0.029 $\pm$ 0.020\\
2448690327240200576 & 0.017 $\pm$ 0.018\\
2448689811844125184 & 0.529 $\pm$ 0.118\\
2448784953959717376 & 0.0454 $\pm$ 0.0126\\
2448783991887042048 & 1.25 $\pm$ 0.14
\enddata
\tablecomments{This table is available in its entirety in machine-readable form.
}
\end{deluxetable}

We now use the criteria in \citet{Bai19b} to select qualified stars in $Gaia$ DR2, and there are 132,739,322 stars left.
The feature space constructed with ten input parameters is applied to regress their {\EBR}, and
the result is listed in Table \ref{Table2}.

\begin{figure}
   \centering
   \includegraphics[width=0.5\textwidth]{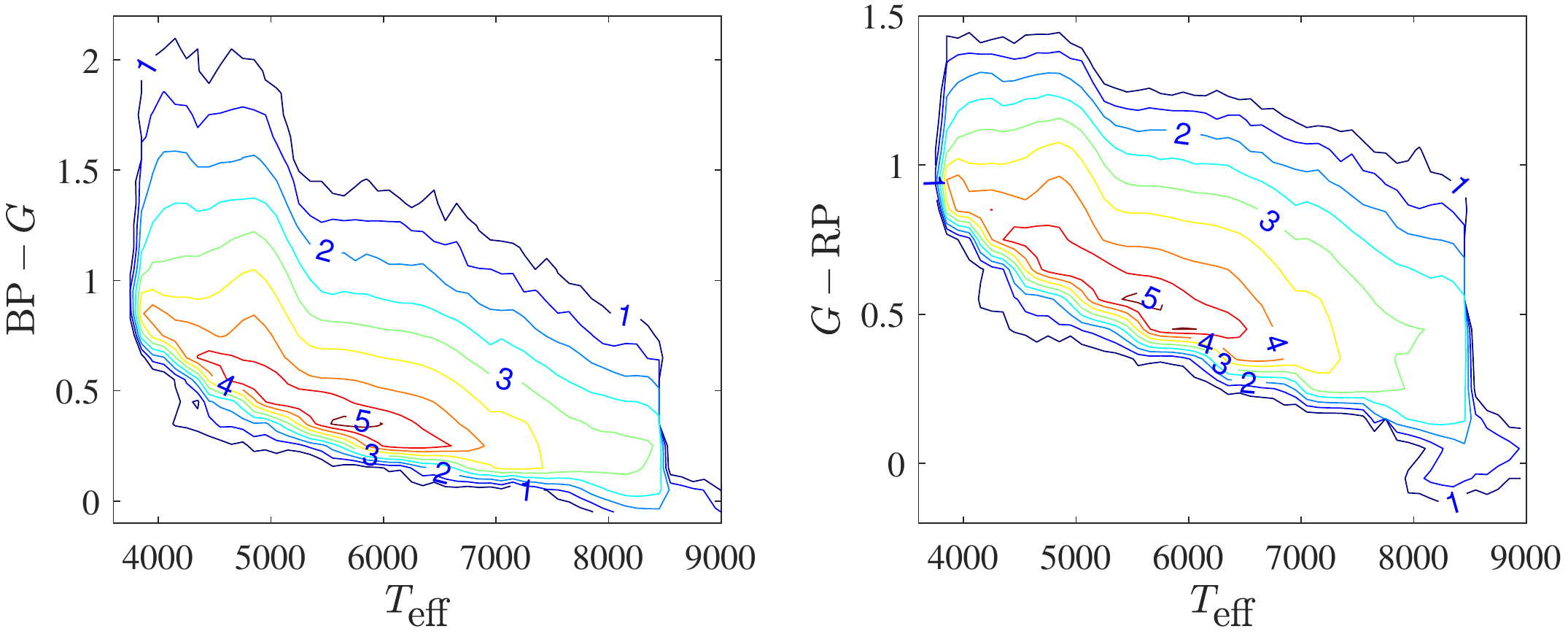}
   \caption{$Gaia$ colors vs. the {\Teff}. The contours are the densities of the stars in our training sample.
   The numbers are the densities in the logarithmic scale.
   \label{Flag}}
\end{figure}

\citet{Bai19b} suggested that external interpolation could regress results with large deviation. We plot two $Gaia$
colors as functions of the temperature in Figure \ref{Flag}.
We use the outmost contour (log Density = 1) to separate 133 million stars into two classes, the stars located outside the contour and
inside the contour. The stars located outside the contour are
externally regressed in these color-temperature spaces.

\begin{figure}
   \centering
   \includegraphics[width=0.5\textwidth]{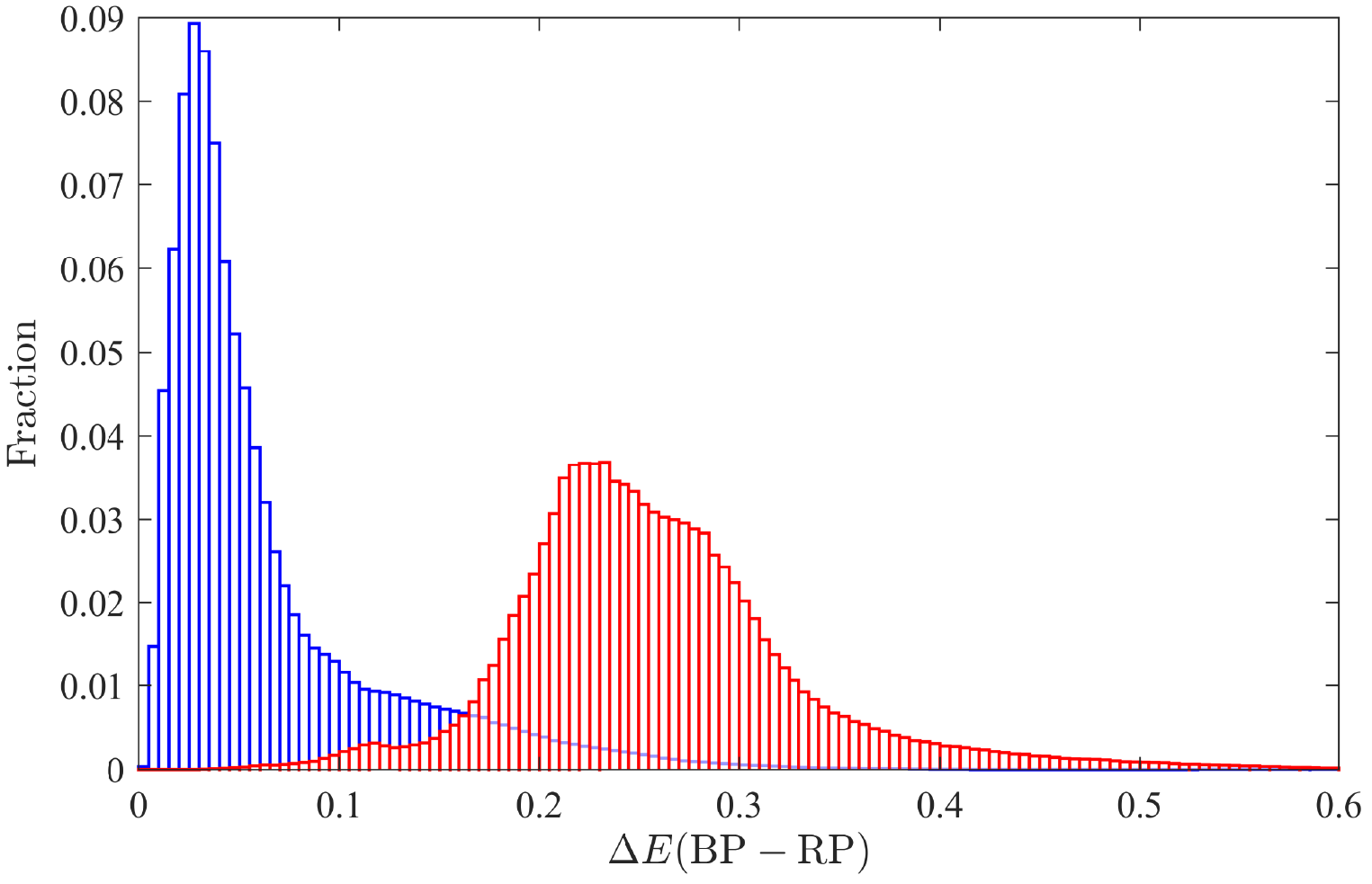}
   \caption{The distribution of the extinction uncertainties. The blue histogram is the stars located inside the outmost
            contour in Figure \ref{Flag}, and the red histogram is the stars located outside the contour.
   \label{HistE}}
\end{figure}

We then present the distribution of the extinction uncertainties in Figure \ref{HistE}, which shows that the stars located outside the
contour tend to have higher deviation, larger than 0.1 mag. This indicates that we could use the uncertainty of the extinction
to discriminate the result from the potential external regression. There are 106,042,018 stars with the uncertainties less
than 0.1 mag.

\begin{figure}
   \centering
   \includegraphics[width=0.5\textwidth]{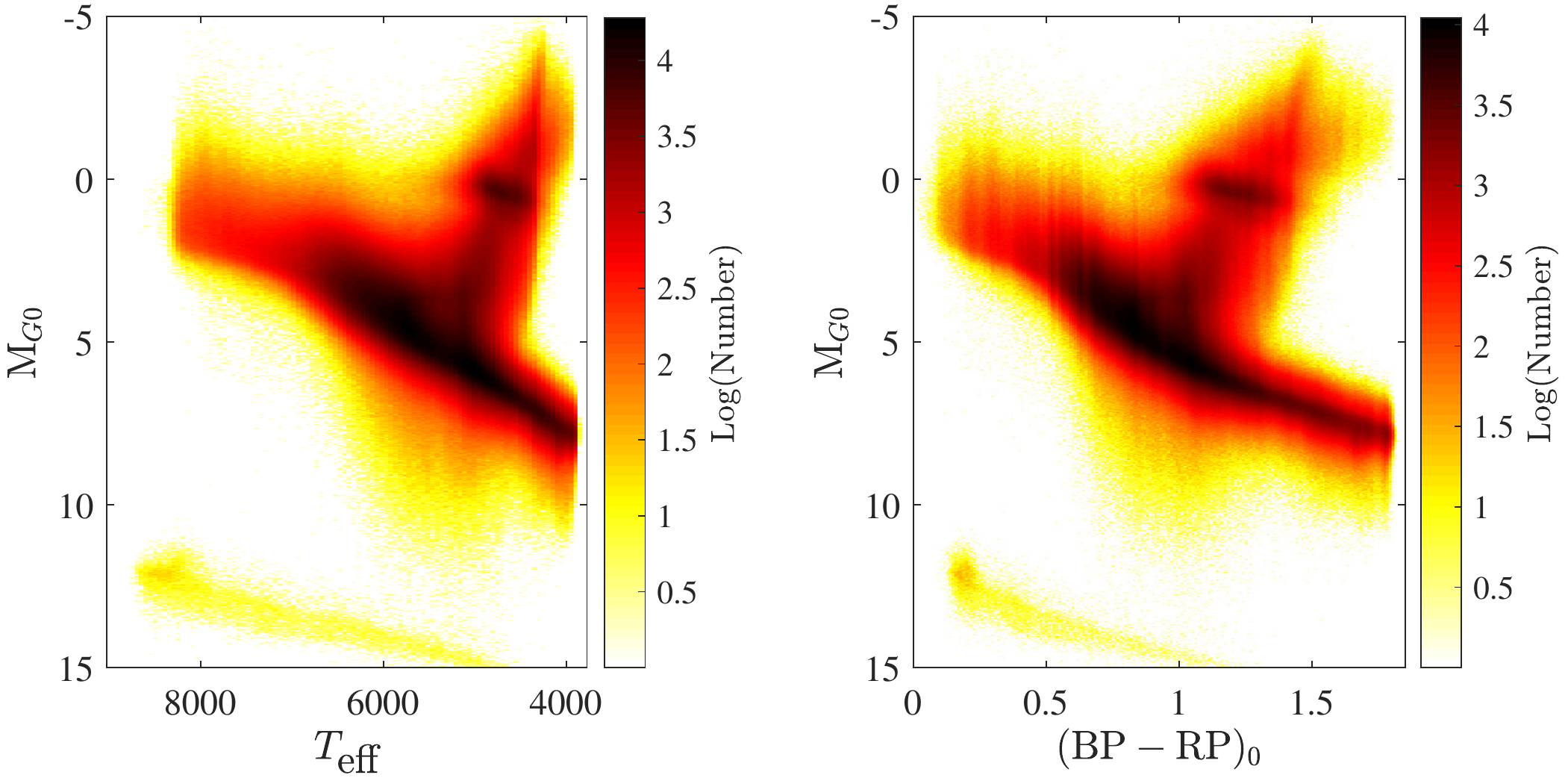}
   \caption{HR-like diagrams for the stars with extinction uncertainties less than 0.1 mag. M$_{G0}$ vs. {\Teff} is in the left panel
            and M$_{G0}$ vs. (BP $-$ RP)$_0$ is in the right panel.
   \label{HRD}}
\end{figure}

The HR-like diagrams are presented in Figure \ref{HRD}. Since the training sample are dominated by the A,
F, G, K stars \citep{Bai19b}, there is no stars blues than ($\textrm{BP} - \textrm{RP})_0$ = 0 or redder than
1.9. We could not find obvious horizontal concentrated lines in the diagram, which is different to the result of \citet{Andrae18}.
The concentrated lines are probably due to the failure of temperature-extinction decoupling and the invalidation of the extinction.
On the other hand, our results are tested within the parameter space covered by the spectroscopic surveys, and externally regressing them to
M or B stars would suffer deviated estimates.

\section{Discussion}
\label{sec:sum}
In this work, we have attempted to regress {\EBR} for 132,739,322 stars in $Gaia$ DR2 using machine
learning algorithm. The regressor is trained with over three million stars in LAMOST, SSPP and APOGEE catalogs.
We adopt stellar temperature, the parameters of Galactic position and two colors to build the regressor.
The performance of the regression is examined with cross validations and a blind test of stars in the RAVE survey,
which indicate that our regressor could predict the stellar extinction with fair accuracy.
In this section we would like to discuss the comparisons with the results in other studies.

\subsection{Photometry-Based Method}
\begin{figure}
   \centering
   \includegraphics[width=0.5\textwidth]{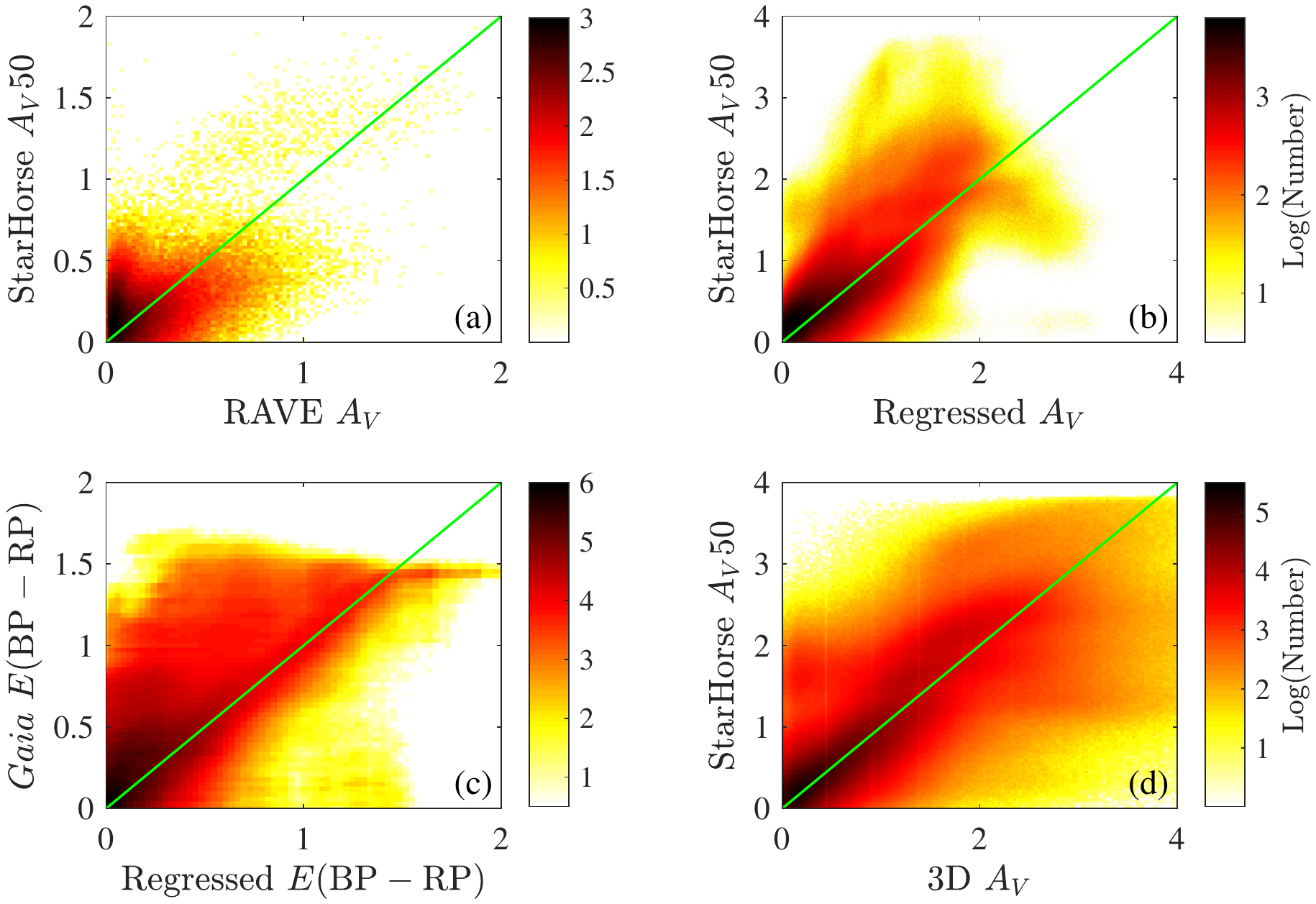}
   \caption{One-to-one correlations. (a) $A_V$50 vs. RAVE $A_V$, (b) $A_V$50 vs. our result, (c) $Gaia$ extinction vs. our result,
            (d) $A_V$50 vs. 3D extinction in \citet{Green19}.
   \label{OneOne}}
\end{figure}

\citet{Anders19} derived the extinction for 265 million stars using the code \textsf{StarHorse}, based on the combination
of $Gaia$ DR2 and the photometric catalogues of Pan-STARRS1, 2MASS and AllWISE. We cross match this catalog
with our result and the RAVE catalog, and present the one-to-one correlations in the upper panels of Figure \ref{OneOne}.
The $A_V$50 stands for the flag-cleaned 50th percentile of the line-of-sight extinction. Here we adopt the coefficient in \citet{Wang19}
to convert {\EBR} to $A_V$. The consistencies are not good between the \textsf{StarHorse} result and those from the spectrum-based methods.
The standard deviation is 0.23 mag for the panel (a) and 0.44 mag for the panel (b),
about 10 times higher than our results of the cross validation and the blind tests.

We present the comparison between our results and the extinction in $Gaia$ DR2 in the panel (c) of Figure \ref{OneOne}. The standard
deviation is about 0.20 mag. There are many stars with extinction over-estimated by $Gaia$ DR2, which is similar to the distribution of the training sample.
There are also some stars located at lower-right area, which aren't shown in Figure \ref{OneOne_Tr}. These stars are probably potential
samples with the external regression, which could not be removed by the color-temperature criteria.

Another popular extinction estimate is a 3D dust map. \citet{Green19} have presented a three-dimensional map of dust reddening,
based on $Gaia$ parallaxes and stellar photometry from Pan-STARRS 1 and 2MASS. We retrieve the extinction of $Gaia$ stars with their
code \textsf{dustmaps}\footnote{https://dustmaps.readthedocs.io/en/latest/}, and match the result to the \textsf{StarHorse}
catalog. The one-to-one correlation is presented in the panel (d) of Figure \ref{OneOne}, which shows large bias with the standard
deviation of 0.40 mag.

As discussed in \citet{Bai19a}, it is not an effective way to describe stellar physical environment only based on stellar photometry,
since the observation conditions and the deviation estimations of different surveys are not consistent. These differences
could produce additional noise, and further propagate to the results. These differences also exist in the spectrum-based surveys,
but a spectrum has about a thousand data points, and it could bring much more information than multi-band photometry.
These differences would become marginal, if we select spectra with high quality and similar resolution.
When the signal-to-noise ratio of the input data goes up, the uncertainty goes down and a more reliable result could be acquired.

Moreover, the performance of the results is algorithm independent. The Bayesian method has been applied in the RAVE
catalog, in the \citet{Wang16a} and \citet{Wang16b}, in the \citet{Anders19} and \citet{Green19}. The spectrum-based results share good consistency,
while photometry-based results have large deviation. The volume and accuracy of the input information have a decisive influence
on the overall performance of the result.

\subsection{Extinction Coefficient}
\citet{Wang19} has presented precise multi-band coefficients for a group of 61,111 red clump stars in the APOGEE survey.
Their coefficient ratio $A_{Ks}$/$A_V$ = 0.078 is lower than the result in our training sample
$\frac{0.2752E\textrm{(BP - RP)}}{2.138E\textrm{(BP - RP)}}$ = 0.129 (Figure \ref{EBV}).
\citet{Dutra02} has built $K$ band extinction maps in the area of two candidate low-extinction windows in the inner Bulge,
and the ratio is 0.118.

It has long been debated whether the infrared extinction law is universal \citep{Wang13,Wang14}.
The dust may be larger in denser regions of the Galaxy, which would lead to a smaller power law index \citep{Li15}.
The APOGEE survey is in the near infrared band that could observe the stars located in the denser regions than the LAMOST
survey which is in the optical wavelength.
These different ratios may imply that the red clump stars of \citet{Wang19} and the APOGEE stars in our training sample are located at a different
regions of the Galaxy. Such difference would slightly differentiate the coefficient in the near infrared.
We check the regions covered by LAMOST, SSPP and APOGEE for the stars in our training sample, and find that most of
them are located at similar regions. Therefore, this difference is not obvious for the three surveys of our training sample.

\begin{acknowledgements}

This work was supported by the National Natural Science Foundation of China (NSFC)
through grants NSFC-11988101/11973054/11933004/11603038 and the National Programs on Key Research and Development
Project (Grant No. 2019YFA0405504 and 2016YFA0400804).
This work presents results from the European Space Agency (ESA) space mission Gaia. Gaia data
are being processed by the Gaia Data Processing and Analysis Consortium (DPAC).
Funding for the DPAC is provided by national institutions, in particular the institutions
participating in the Gaia MultiLateral Agreement (MLA). The Gaia mission website is
\url{https://www.cosmos.esa.int/gaia}. The Gaia archive website is \url{https://archives.esac.esa.int/gaia}.

The Guoshoujing Telescope (the Large Sky Area Multi-Object
Fiber Spectroscopic Telescope, LAMOST) is a National Major
Scientific Project which is built by the Chinese Academy of
Sciences, funded by the National Development and Reform Commission,
and operated and managed by the National Astronomical Observatories,
Chinese Academy of Sciences.

Funding for the Sloan Digital Sky Survey IV has been provided by the Alfred P.
Sloan Foundation, the U.S. Department of Energy Office of Science, and the
Participating Institutions. SDSS-IV acknowledges
support and resources from the Center for High-Performance Computing at
the University of Utah. The SDSS web site is \url{http://www.sdss.org/}.

SDSS-IV is managed by the Astrophysical Research Consortium for the
Participating Institutions of the SDSS Collaboration including the
Brazilian Participation Group, the Carnegie Institution for Science,
Carnegie Mellon University, the Chilean Participation Group, the French
Participation Group, Harvard-Smithsonian Center for Astrophysics,
Instituto de Astrof\'isica de Canarias, The Johns Hopkins University,
Kavli Institute for the Physics and Mathematics of the Universe (IPMU) /
University of Tokyo, Lawrence Berkeley National Laboratory,
Leibniz Institut f\"ur Astrophysik Potsdam (AIP),
Max-Planck-Institut f\"ur Astronomie (MPIA Heidelberg),
Max-Planck-Institut f\"ur Astrophysik (MPA Garching),
Max-Planck-Institut f\"ur Extraterrestrische Physik (MPE),
National Astronomical Observatories of China, New Mexico State University,
New York University, University of Notre Dame,
Observat\'ario Nacional / MCTI, The Ohio State University,
Pennsylvania State University, Shanghai Astronomical Observatory,
United Kingdom Participation Group,
Universidad Nacional Aut\'onoma de M\'exico, University of Arizona,
University of Colorado Boulder, University of Oxford, University of Portsmouth,
University of Utah, University of Virginia, University of Washington, University of Wisconsin,
Vanderbilt University, and Yale University.
\end{acknowledgements}

\end{document}